\documentclass[12pt,a4paper]{article}

\usepackage{amsmath,amsthm,amstext,amscd,amssymb,euscript,mathrsfs}
\usepackage{calrsfs}
\usepackage{epsfig}

\newcommand{\Z}{\mathbb Z}
\newcommand{\R}{\mathbb R}
\newcommand{\N}{\mathbb N}

\newcommand{\E}{\mathbb E}

\renewcommand{\Pr}{\mathbb P}

\newcommand{\epsi}{\ensuremath{\epsilon}}
\newcommand{\La}{\ensuremath{\Lambda}}

\def\1{{\mathchoice {\rm 1\mskip-4mu l} {\rm 1\mskip-4mu l}
{\rm 1\mskip-4.5mu l} {\rm 1\mskip-5mu l}}}

\newtheorem{theorem}{{\small T}{\scriptsize HEOREM}}[section]
\newtheorem{corollary}{{\bf{\small C}{\scriptsize OROLLARY}}}[section]
\newtheorem{proposition}{{\bf{\small P}{\scriptsize ROPOSITION}}}[section]
\newtheorem{lemma}{{\bf{\small L}{\scriptsize EMMA}}}[section]
\newtheorem{remark}{{\bf{\small R}{\scriptsize EMARK}}}[section]
\newtheorem{definition}{{\bf{\small D}{\scriptsize EFINITION}}}[section]

\renewenvironment{proof}[1]
{\noindent{{\bf{\small{ P}{\scriptsize ROOF}}}.}\hspace{0.1cm} #1} {$\;\qed$\newline}

\newcommand{\beq}{\begin{eqnarray}}
\newcommand{\eeq}{\end{eqnarray}}

\newcommand{\ba}{\begin{align*}}
\newcommand{\ea}{\end{align*}}

\newcommand{\be}{\begin{equation}}
\newcommand{\ee}{\end{equation}}

\newcommand{\bl}{\begin{lemma}}
\newcommand{\el}{\end{lemma}}

\newcommand{\br}{\begin{remark}}
\newcommand{\er}{\end{remark}}

\newcommand{\bt}{\begin{theorem}}
\newcommand{\et}{\end{theorem}}

\newcommand{\bd}{\begin{definition}}
\newcommand{\ed}{\end{definition}}

\newcommand{\bp}{\begin{proposition}}
\newcommand{\ep}{\end{proposition}}

\newcommand{\bc}{\begin{corollary}}
\newcommand{\ec}{\end{corollary}}

\newcommand{\bpr}{\begin{proof}}
\newcommand{\epr}{\end{proof}}

\newcommand{\bi}{\begin{itemize}}
\newcommand{\ei}{\end{itemize}}

\newcommand{\ben}{\begin{enumerate}}
\newcommand{\een}{\end{enumerate}}

\begin{document}
\title{Condensation in the inclusion process and related models}
\author{Stefan Grosskinsky$^{\textup{{\tiny(a)}}}$,
Frank Redig$^{\textup{{\tiny(b)}}}$, Kiamars Vafayi$^{\textup{{\tiny(c)}}}$\\
{\small $^{\textup{(a)}}$ Mathematics Institute, University of Warwick, Coventry CV4 7AL, UK}\\
{\small $^{\textup{b)}}$ IMAPP, Radboud University Nijmegen}\\
{\small Heyendaalse weg 135,
6525 AJ Nijmegen, The Netherlands}\\
{\small $^{\textup{(c)}}$ Mathematisch Instituut Universiteit Leiden}\\
{\small Niels Bohrweg 1, 2333 CA Leiden, The Netherlands}\\
}

\maketitle

\begin{abstract}
We study condensation in several particle systems related to the inclusion process. For an asymmetric one-dimensional version with closed boundary conditions and drift to the right, we show that all but a finite number of particles condense on the right-most site. This is extended to a general result for independent random variables
with different tails, where condensation occurs for the index (site) with the heaviest tail, generalizing also previous results for zero-range processes. For inclusion processes with homogeneous stationary measures we establish condensation in the limit of vanishing diffusion strength in the dynamics, and give several details about how the limit is approached for finite and infinite systems.
Finally, we consider a continuous model dual to the inclusion process, the so-called Brownian energy process, and prove similar condensation results.

\bigskip


\end{abstract}


\section{Introduction}
In \cite{gkrv}, \cite{grv}, an interacting particle system was introduced,
where particles perform random
walks and interact by ``inclusion'', i.e., every particle at site $i$
can attract particles from a site $j$ to its site at rate $p(i,j)=p(j,i)$. This particle system,
the so-called symmetric inclusion process (SIP), is ``exactly solvable" by self-duality, and
its ergodic stationary measures are products of discrete gamma distributions, indexed by the density. The inclusion process also turns out to be dual to a system of interacting diffusions, the
so-called Brownian energy process (BEP).
More details on duality, self-duality, and the precise relations
between SIP and BEP can be found in \cite{gkrv}. In the present paper we only need the explicit form of the stationary measures
of these models.


We prove existence of stationary product measures for inclusion processes under rather general conditions, in analogy to classical results for exclusion processes \cite{liggett}.
We introduce asymmetric versions of the SIP and the BEP,  for simplicity focusing on a one-dimensional context with $N$ sites and closed boundary conditions. In this case both models have spatially inhomogeneous product measures as reversible measures (to be compared with the blocking measure of
the asymmetric exclusion process).
Conditioning on $K$ particles in the system (resp.\ total energy $E$),
we prove that that in the limit $K\to\infty$ ``almost all" the particles (resp.\ all the energy)
are concentrated on a single site, where the marginal of the reversible measure has the heaviest tail. The other
sites contain a finite number of particles (resp.\ finite amount of energy).

We further study condensation in inclusion processes with spatially homogeneous stationary measures, with the SIP as the main example. The strength of the diffusive part of the dynamics in comparison to the attraction is controlled by a system parameter $m>0$. For fixed particle density $\rho$ we study the limit $m\to 0$ where attraction dominates, and show that the single-site marginals converge to Dirac measures concentrated on zero mass. This  corresponds to the fact that a typical configuration consists of rare piles of typical size $2\rho /m$ separated by empty sites. The distribution of pile sizes approaches a power law with exponent $-1$ and becomes degenerate in the limit $m\to 0$. This leads to a breakdown of the usual law of large numbers which we illustrate in detail.

Our results for the asymmetric case also cover condensation phenomena in zero-range processes, which have attracted a lot of recent research interest \cite{evans00,evansetal05}. For inhomogeneous systems, these have been studied before mainly in the context of a quenched disorder in the jump rates, which have to be non-decreasing functions of the number of particles \cite{krug,landim96,andjeletal00,ferrarisisko}. For such systems, the use of coupling techniques allowed in special cases to also obtain results on the dynamics of condensation. In contrast, our results cover only the stationary behaviour but apply to a much larger class of jump rates with essentially no restriction. The widely studied condensation in spatially homogeneous zero-range processes \cite{grossketal03,godreche,sisko,armendarizetal09,beltran} has a somewhat different origin than our homogeneous results for the SIP. This is discussed in detail at the end of Section 4.2.

In the next section we describe the inclusion process and its stationary measures. In Sections 3 and 4 we study condensation in the asymmetric and spatially homogeneous case, and discuss extensions and relations to zero-range processes. In Section 5 we introduce the asymmetric Brownian energy process and discuss condensation in an example of a system with continuous state space.

\section{Inclusion processes}

The inclusion process on a general discrete set $\La$ has state space $\Omega= \N^{\La}$ and we denote a configuration by $\eta =(\eta_i :i\in\La )$ where $\eta_i$ is interpreted as the number of particles at site $i\in\La$. The dynamics is defined by the generator defined on the core of local functions $f:\Omega\to\R$:
\be\label{ipgen}
L f(\eta)=\sum_{i,j\in\Lambda} p(i,j) \eta_i \left( \frac{m}{2} +\eta_{j}\right)
\big( f(\eta^{i,j})-f(\eta)\big)\ ,
\ee
where $\eta^{i,j}$ is the configuration
obtained from $\eta$ by removing a particle from site $i$ and
putting it to $j$.
The $p(i,j)\geq 0$ are jump rates of an irreducible random walk on $\Lambda$ with $p(i,i)=0$, and the parameter $m>0$ determines the rate of diffusion of the particles as compared to the aggregation part given by the product $\eta_i \eta_j$. We also assume the $p(i,j)$ to be uniformly bounded and of finite range, i.e. there exist $C,R>0$ such that
\be
\sup_{i,j\in\Lambda} p(i,j)<C\quad\mbox{and}\quad \left|\big\{ j\in\Lambda\, :\, p(i,j)>0\big\}\right| <R\ \mbox{for all }i\in\Lambda\ .
\ee
This ensures that the dynamics is well defined even on infinite lattices (for a large class of 'reasonable' initial conditions) and contains all generic examples we are interested in, such as nearest-neighbour hopping on regular lattices.\\
If the $p(i,j)$ are symmetric the inclusion process is also called symmetric (SIP), otherwise asymmetric (ASIP).

\subsection{Stationary product measures}

For $\phi\geq 0$ and $\lambda_i >0$, $i\in\Lambda$, define the product probability measure
\be\label{ippro}
\nu_\phi (d\eta)= \otimes_{i\in\Lambda} \nu^i_\phi (d\eta_i)\ ,
\ee
where the marginals $\nu^i$ are probability measures on $\N$ given by
\be\label{ipmar}
\nu_\phi^i (n)= \left(z_i(\phi)\right)^{-1} \lambda_i^n \phi^n
\frac{\Gamma\left(\frac{m}2+n\right)}{n!\Gamma\left(\frac{m}2\right)}
\ee
with the normalizing constant
\be\label{ipz}
z_i(\phi)= \sum_{n=0}^\infty \lambda_i^n \phi^n
\frac{\Gamma\left(\frac{m}2+n\right)}{n!\Gamma\left(\frac{m}2\right)}
= \left(1-\lambda_i \phi\right)^{-m/2}\ .
\ee
The parameter $\phi\geq 0$ is called fugacity and controls the particle density, which is invariant under the time evolution.

\bt\label{thstat}
For all $\phi< \phi_c :=\big(\sup_{i\in\Lambda}\lambda_i \big)^{-1}$, $\nu_\phi$ is a stationary measure for the inclusion process with generator (\ref{ipgen}), provided that one of the following conditions holds:
\begin{itemize}
    \item[a)] The $p(i,j)$ are doubly stochastic modulo a constant, i.e.
    \be\label{dstoch}
    \sum_{j\in\Lambda} \big( p(i,j)-p(j,k)\big) =0\quad\mbox{for all }i,k\in\Lambda\ ,
    \ee
    and $\lambda_i =1$ for all $i\in\Lambda$.
    \item[b)] The $\lambda_i$ are reversible w.r.t. the $p(i,j)$, i.e.
    \be\label{lrev}
    \lambda_i p(i,j)=\lambda_j p(j,i)\quad\mbox{for all }i,j\in\Lambda\ ,
    \ee
    and in that case $\nu_\phi$ is also a reversible measure.
\end{itemize}
\et

This is in direct analogy with well-known results for stationary measures for exclusion processes (see e.g. \cite{liggett}, Thm VIII.2.1). In both cases, the $\lambda_i$ are special harmonic functions solving
\be\label{lharm}
\sum_{j\in\Lambda} \big(\lambda_i p(i,j)-\lambda_j p(j,i)\big) =0\quad\mbox{for all }i\in\Lambda\ ,
\ee
i.e. they provide a (not necessarily normalized) stationary distribution for the underlying random walk of a single particle. For the above product measures to be stationary, the $p(i,j)$ have to be such that they admit a constant solution (first case) or a detailed balance solution (second case). It is not clear at this point whether these conditions are really necessary for the existence of stationary product measures in general. Note also that on infinite lattices $\phi_c =0$ is possible. But for finite $\Lambda$ (which we mainly focus on in this paper), Theorem \ref{thstat} guarantees the existence of a family of stationary measures.\\

\bpr
We have to show for expected values w.r.t. $\nu_\phi$ that
\be\label{lzero}
\nu_\phi (Lf)=\sum_{\eta\in\Omega} \sum_{i,j\in\Lambda} p(i,j) \eta_i \left( \frac{m}{2} +\eta_{j}\right) (f(\eta^{i,j})-f(\eta))\nu_\phi (\eta )=0
\ee
for all local functions $f$. For fixed $i,j$ we get after a change of variable
\beq
\lefteqn{\sum_{\eta\in\Omega} p(i,j) \eta_i \left( \frac{m}{2} +\eta_{j}\right) f(\eta^{i,j}) \nu_\phi (\eta ) }\nonumber\\
& &\qquad =\sum_{\eta\in\Omega} p(i,j) (\eta_i +1)\left( \frac{m}{2} +\eta_{j} -1\right) f(\eta ) \nu_\phi (\eta^{j,i} )\ .\nonumber
\eeq
The form (\ref{ipmar}) of the marginals implies that for all $i\in\Lambda$ and $k\geq 0$
\[
\frac{\nu_\phi^i (k+1)}{\nu_\phi^i (k)}= \phi\, \frac{m+2k}{2(k+1)} \,\lambda_i \ .
\]
Thus we get for each fixed pair $i,j\in\Lambda$
\[
\nu_\phi^i (n{+}1)\,\nu_\phi^j (k{-}1)\, (n+1)\left( \frac{m}{2} +k -1\right) =\nu_\phi^i (n)\,\nu_\phi^j (k)\, k\left( \frac{m}{2} +n\right)\frac{\lambda_i}{\lambda_j}
\]
for all $n\geq 0$ and $k\geq 1$. It is easy to check that boundary terms in the sums vanish consistently, and we do not consider them in the following. Plugging this into (\ref{lzero}) we get
\be\label{final1}
\nu_\phi (Lf)=\sum_{\eta\in\Omega} f(\eta )\nu_\phi (\eta )\sum_{i,j\in\Lambda} p(i,j) \left( \eta_j \left( \frac{m}{2} {+}\eta_{i}\right)\frac{\lambda_i}{\lambda_j}{-} \eta_i \left( \frac{m}{2} {+}\eta_{j}\right)\right)\ ,
\ee
and exchanging the summation variables $i\leftrightarrow j$ in the first part of the sum leads to
\be\label{final2}
\nu_\phi (Lf)=\sum_{\eta\in\Omega} f(\eta )\nu_\phi (\eta )\sum_{i,j\in\Lambda} \frac{\eta_i \left( m/2 +\eta_{j}\right)}{\lambda_i} \big( p(j,i)\lambda_j -p(i,j)\lambda_i\big)\ .
\ee
This clearly vanishes under the reversibility condition (\ref{lrev}) which implies stationarity under assumption b). In analogy to (\ref{final1}) we can derive
\beq
\nu_\phi (gLf)&=&\sum_{\eta\in\Omega} f(\eta )\nu_\phi (\eta )\nonumber\\
& &\quad\sum_{i,j\in\Lambda} p(i,j) \left( \eta_j \left( \frac{m}{2} {+}\eta_{i}\right)\frac{\lambda_i}{\lambda_j} g(\eta^{j,i})- \eta_i \left( \frac{m}{2} {+}\eta_{j}\right) g(\eta )\right)\ ,\nonumber
\eeq
and after using (\ref{lrev}) and the exchange of summation variables this implies
\beq
\nu_\phi (gLf)&=&\sum_{\eta\in\Omega} f(\eta )\nu_\phi (\eta )\sum_{i,j\in\Lambda} p(i,j) \eta_i \left( \frac{m}{2} {+}\eta_{j}\right) \big(g(\eta^{i,j})- g(\eta )\big)\nonumber\\
& =&\nu_\phi (fLg)\ ,\nonumber
\eeq
so $\nu_\phi$ is also reversible.\\
Assuming a), the $\lambda_i$ and $\lambda_j$ in (\ref{final2}) cancel, and we write the linear (diffusive) part as
\[
\sum_{i\in\Lambda} \eta_i \frac{m}{2} \sum_{j\in\Lambda} \left( p(j,i) -p(i,j)\right) =0\ ,
\]
which vanishes due to (\ref{dstoch}). For the quadratic aggregation part we get
\[
\sum_{i,j\in\Lambda} \eta_i \eta_j \left( p(j,i) -p(i,j)\right) =\sum_{i,j\in\Lambda} \eta_i \eta_j \left( p(j,i) -p(j,i)\right) =0\ ,
\]
by another exchange of the summation variables in the second part, using that $\eta_i \eta_j$ is symmetric under $i\leftrightarrow j$.
\epr

\subsection{Canonical measures for finite systems}

Consider a finite lattice $\Lambda_N$ of size $N$ with corresponding state space $\Omega_N =\N^{\Lambda_N}$. Starting with a fixed number of $K$ particles, the inclusion process with generator $L_N$ as given in (\ref{ipgen}) is an irreducible continuous-time Markov chain on the finite set
\be\label{ak}
A_K =\left\{\eta\in\Omega_N:\sum_{i=1}^N \eta_i =K\right\}\ ,
\ee
and has a unique stationary measure, which we denote by $\mu^K$.

By conservation of the number of particles, the conditional measure
\[
\nu_\phi \left( d\eta \left|\sum_{i=1}^N\eta_i=K\right.\right)
\]
is also invariant. Indeed for $f:\Omega_N\to\R$ we have
\beq\label{mustat}
\int L_N f (\eta) \nu_\phi \left( d\eta \big|A_K\right)
&=&
\frac{\int L_N f(\eta) 1_{A_K} \nu_\phi (d\eta)}
{\nu_\phi (A_K)}
\nonumber\\
&=&
\frac{\int  f(\eta) \left(L_N^* \left(1_{A_K}\right)\right)(\eta) \nu_\phi (d\eta)}
{\nu_\phi (A_K)} =0\ ,
\eeq
since it is easy to see that with the generator $L_N$ also its adjoint $L_N^*$ conserves the number of particles. In the case of reversible measures $\nu_\phi$, $L_N$ is self-adjoint and there is nothing to check.
By uniqueness of the stationary measure, we thus have
\beq
\nu_\phi (.\, |A_K)=\mu^K
\eeq
for all $\phi <\phi_c$ and $K\in \N$. So the conditioned product measures are actually independent of $\phi$, and this connection provides an explicit form for the canonical measures $\mu^K$.

\section{Condensation in the ASIP}

A generic situation where Theorem \ref{thstat} gives rise to spatially inhomogeneous reversible measures is a one-dimensional lattice $\Lambda_N =\{ 1,\ldots ,N\}$ with an underlying asymmetric nearest-neighbour walk. We consider the ASIP with generator
\beq\label{asipgen}
L_N f(\eta)&=&\sum_{i=1}^{N-1} p \eta_i \left( \frac{m}{2} +\eta_{i+1}\right)
(f(\eta^{i,i+1})-f(\eta))
\nonumber\\
& &+ \sum_{i=1}^{N-1} q \eta_{i+1} \left( \frac{m}{2} +\eta_{i}\right)
(f(\eta^{i+1,i})-f(\eta))
\eeq
where $p>q>0$. 
In this case $\lambda_i =(p/q)^i$ fulfills condition (\ref{lrev}) in Theorem \ref{thstat}.
We will now proceed towards showing that in the limit
$K\to\infty$, under the canonical measure $\mu^K$, the typical
situation will be that all but a finite number of particles condenses at the right site $i=N$, whereas the other sites contain a number
of particles distributed according to $\nu^i_{\phi_c}$.

At first sight one could be tempted to think that this is just
a consequence of the asymmetry: particles are pushed to the
right. This is, however, not the case. If we
consider independent random walkers, moving
at rate $p$ to the right and $q$ to the left, then
the reversible profile measures are Poissonian and given
by $\otimes_{i=1}^N \nu^i_\phi(d\eta_i)$ with
\[
\nu^i_\phi (n) = \frac{1}{z_i (\phi)}\left(\frac{p}{q}\right)^{ni}\frac{\phi^n}{n!}
\]
with a normalizing constant $z_i (\phi)= e^{\phi (p/q)^i}$ which
is now finite for all values of $\phi$. As a consequence,
no condensation happens: if we condition on having $K$ particles,
and let $K$ tend to infinity, all sites will carry a diverging number of particles. The condensation phenomenon is thus a combination
of the asymmetry, together with the attractive interaction between
the particles in the inclusion process. Indeed, it is the interaction
which is responsible for the existence of a finite critical $\phi_c$.

\subsection{Condensation}

Before we formulate the main result of this Section, we recall the marginals $\nu^i_\phi$ for the ASIP
\be\label{generalform}
\nu^i_\phi (n)= \frac{1}{z_i (\phi)} \,\phi^n \lambda_i^n w_i (n)\ ,
\ee
where we have now
\be\label{lambda}
\lambda_i = (p/q)^i
\ee
and write
\be\label{weights}
w_i (n)=\frac{ \Gamma \left(n+\frac{m}{2}\right)}{n!\Gamma (n)}\ .
\ee
In the present case $w_i$ does not dependend on $i$, but
in generalizations explained below we will allow explicit dependence
on $i$. The weights $w_i (n)$ have the asymptotic behavior
\be\label{asweight}
w_i (n) \sim n^{\frac{m}{2}-1}
\ee
where $a_n\sim b_n$ means that $a_n/b_n$ converges to a strictly positive constant.

We remind that the normalizing constants are
\be\label{asipz}
z_i(\phi)
= \left(1-\lambda_i \phi\right)^{-m/2} =\left(1-\left(\frac{p}{q}\right)^i \phi\right)^{-m/2}\ .
\ee
Therefore, in the context of Theorem \ref{thstat} we have $\lambda_1 <\lambda_2<\ldots <\lambda_N$, $\phi_c = 1/\lambda_N$,
$z_i (\phi_c) <\infty$ for all $1\leq i\leq N-1$, and $z_N (\phi) <\infty$ for
all $\phi <\phi_c$. We then have the following result.

\bt\label{condt}
\begin{itemize}
\item[a)] In the limit $K\to\infty$, $\eta_1,\ldots, \eta_{N-1}$ are asymptotically independent and converge in distribution to the critical product measure, i.e. for all $n_1 ,\ldots ,n_{N-1}\in\N$
\be\label{kolibri}
\mu^K\left(\eta_1=n_1 ,\ldots ,\eta_{N{-}1}=n_{n{-}1}\right)\to
\nu_{\phi_c}^1 (n_1)\cdots \nu_{\phi_c}^{N{-}1} (n_{N{-}1})
\ee
where $\phi_c= 1/\lambda_N =(q/p)^N$.
\item[b)] In the limit $K\to\infty$, the right edge contains ``almost
all" particles, i.e., for all $\delta\in (0,1)$
\be\label{kraai}
\mu^K \big(\eta_N \leq (1-\delta)K\big)\to 0\ ,
\ee
and we have a strong law of large numbers,\quad $\eta_N /K\to 1\quad a.s.$\ .
\end{itemize}
\et

\bpr
We use that $\mu^K= \nu_\phi (.\, |A_K)$ and write for $\La'\subseteq\Lambda_N$
\be\label{part}
Z(\La',K)= \sum_{\{n_i,i\in\La':\sum_{i\in\La'}n_i=K\}} \prod_{i\in\La'} w_i (n_i)\lambda_i^{n_i}\ .
\ee
We then have
\beq\label{bombom}
\lefteqn{
\mu^K\left(\eta_1=n_1 ,\ldots ,\eta_{N{-}1}=n_{n{-}1}\right) =}\nonumber\\
& &= \frac{1}{Z(\Lambda_N ,K)}\ w_N \left( K-\sum_{i=1}^{N-1} n_i \right)\,\lambda_N^{K-\sum_{i=1}^{N-1} n_i} \prod_{i=1}^{N-1} w_i (n_i)\,\lambda_i^{n_i} \ .
\eeq
We first prove that
\be\label{bambam}
\lim_{K\to\infty} \frac{Z(\La_N, K)}{\lambda_N^K w_N (K)} = \prod_{i=1}^{N-1} z_i (\phi_c)\ .
\ee
To see this, we choose an appropriate order of summation,
\beq\label{boemboem}
Z(\La_N, K) &=&
\sum_{n_1=0}^K \sum_{n_2= 0}^{K-n_1}\ldots\sum_{n_{N-1}=0}^{K-(n_1+\ldots n_{N-2})}\nonumber\\
& &\qquad\quad\left(\prod_{j=1}^{N-1} w_j (n_j)\lambda_j^{n_j}\right) w_N\left( K-\sum_{j=1}^{N-1} n_j\right)
\lambda_N^{K-\sum_{j=1}^{N-1} n_j}
\nonumber\\
&=&
\lambda_N^K w_N(K)\sum_{n_1=0}^\infty\ldots \sum_{n_{N-1}=0}^\infty\nonumber\\
& &\qquad\qquad\Psi_K (n_1,\ldots, n_{N-1}) \prod_{j=1}^{N-1} w_j (n_j)
\left(\frac{\lambda_j}{\lambda_N}\right)^{n_j}
\eeq
with
\be\label{blop}
\Psi_K (\ldots)=
\frac{w_N \left(K{-}\sum_{j=1}^{N-1} n_j\right)}{w_N (K)}\, 1_{n_1\leq K}\cdots 1_{n_{N{-}1}\leq K{-}n_1{-}..{-}n_{N{-}2}}\ .
\ee
We see from \eqref{weights} that
$\Psi_K\leq 1$. Therefore, by dominated
convergence, using that $\phi_c= \lambda_N^{-1}$ and
\[
z_i (\phi_c) =\sum_{n=0}^\infty \phi_c^n \, w_i (n)\, \lambda_i^n = \sum_{n=0}^\infty w_i (n)
\left(\frac{\lambda_i}{\lambda_N}\right)^n <\infty\ ,
\]
we obtain \eqref{bambam}.
Combining 
\eqref{bombom} and \eqref{bambam} with the fact that
\be\label{een}
\lim_{K\to\infty} \frac{w_N (K-n)}{w_N(K)}=1\quad\mbox{for all }n\in\N
\ee
(which follows immediately from \eqref{asweight}),
yields item a) of Theorem \ref{condt}.

To prove item b), we start with
\[
\mu^K \big(\eta_N \leq (1-\delta)K\big)
=
\frac{\sum_{n\leq (1-\delta)K} w_N (n) \lambda_N^n Z(\La_N\setminus\{ N\}, K-n)}
{Z(\La_N, K)}
\]
and estimate, for $n\leq (1-\delta)K$ and a small enough $\epsi'>0$ to be chosen below:
\beq\label{coala}
Z(\La_N\setminus \{ N\}, K-n)
&\leq &\left(\lambda_{N-1}(1+\epsi')\right)^{K-n}\nonumber\\
& &\sum_{n_1=0}^{K-n}\ldots \sum_{n_{N-2}=0}^{K-n-(n_1+\ldots+n_{N-3})}
w_{N-1}\left( K-n-\sum_{j=1}^{N-2}n_j\right)\nonumber\\
& &\qquad\left(\prod_{j=1}^{N-2} w_j (n_j) \left(\frac{\lambda_j}{\lambda_{N-1}(1+\epsi')}\right)^{n_j}\right)
\nonumber\\
&\leq &
C \left(\lambda_{N-1}(1+\epsi')\right)^{K-n} (1+\epsi)^{K}\ .
\eeq
Here we have used that (cf.\ \eqref{asweight})
\be\label{twee}
w_{N-1} \left(K-n-\sum_{j=1}^{N-2} n_j\right)\leq C (1+\epsi)^{K}
\ee
for some $\epsi >0$ to be chosen below, and the fact that the remaining
sums in the RHS of \eqref{coala} converge to a finite value
as $K\to\infty$.
By \eqref{bambam} $Z(\La_N, K)$ is bounded below by
$C'\lambda_N^K w_N (K)$ for $K$ large enough.
This then gives
\[
\mu^K \big(\eta_N\leq (1{-}\delta) K\big) \leq C''
\left(\sum_{n\leq (1-\delta)K}
\frac{w_N (n)}{w_N (K)}\right)
\left(\frac{ (1{+}\epsi)^{1/\delta}(1{+}\epsi')\lambda_{N-1}}{\lambda_N}\right)^{\delta K}\ ,
\]
since for the summation indices $K-n\geq \delta K$.
Choosing $\epsi, \epsi'>0$ small enough such that
\[
0<\frac{(1+\epsi)^{1/\delta}(1+\epsi')\lambda_{N-1}}{\lambda_N} < q<1
\]
and using that $\frac{w_N (n)}{w_N (K)}\leq 1$\ ,
we obtain
\be\label{drie}
\mu^K (\eta_N\leq (1-\delta) K)\leq C''q^{\delta K}\ .
\ee
Choosing $\delta =\delta_K =1/\sqrt{K} \to 0$, we get a summable bound on the right-hand side.
Since by definition $\eta_N\leq K\ a.s.$ under the measure $\mu^K$, this implies almost sure convergence and the strong law $\eta_N /K \to 1$ by Borel-Cantelli.
\epr

\subsection{Generalizations}

Notice that in the proof of Theorem \ref{condt}
we did not use the specific form of $w_i$ and $\lambda_i$.
Therefore, the same proof shows a condensation phenomenon
for a general family of independent random variables $\eta_1 ,\ldots ,\eta_N$
with
\[
\Pr (\eta_i =n )=\frac{1}{z_i (\phi)} \, w_i (n)\lambda_i^n \phi^n
\]
under the following hypotheses on the $w_i, \lambda_i$:
\begin{itemize}
\item[a)] The $\lambda_i$ satisfy
\be\label{lambdacond}
\lambda_N>\max_{i=1}^{N-1}\lambda_{i}\ ,
\ee
\item[b)] the weights $w_i(n)$ are subexponential in the following sense
\be\label{subex}
\lim_{n\to\infty} \frac{w_i(n+1)}{w_i (n)} =1
\ee
for all $1\leq i\leq N$.
\end{itemize}
From \eqref{subex}, \eqref{een} follows directly, and it further implies
that for all $\alpha >0$ there exists $C_\alpha >0$ such that for all $n\geq 0$
\[
C_\alpha^{-1} e^{-\alpha n}\leq w_i (n) \leq C_\alpha e^{\alpha n} \ .
\]
From this bound we conclude that for all $\beta >0$, there
exists $C_\beta >0$ such that for all $n,l\geq 0$,
\[
C_\beta^{-1} e^{-\beta (n+l)}\leq \frac{w_i (n)}{w_i (l)} \leq C_\beta e^{\beta (n+l)}\ .
\]
This is all we need in the dominated convergence argument
to bound $\Psi_K$ of \eqref{blop}, and to conclude
\eqref{twee}, \eqref{drie}.
Therefore, under the assumptions
a), b) we conclude the statement of Theorem
\ref{condt} with
\[
\mu^K = \Pr\left(\cdot \left| \sum_{i=1}^N \eta_i =K\right.\right)\ .
\]

\noindent\textbf{Example: Zero-range processes}\\
Consider a general zero-range process on $\Omega_N = \N^{\{ 1,\ldots, N\}}$ with
generator
\be\label{zerrange}
L_N f(\eta) = \sum_{i,j} p(i,j)\, g_i (\eta_i) \left( f(\eta^{i,j}) - f(\eta)\right)\ ,
\ee
where $p(i,j)$ are rates of an irreducible continuous-time random
walk on $\{1,\ldots, N\}$ and where
$g_i : \N\to [0,\infty)$ with $g_i (n)=0$  if and
only if $n=0$.
Moreover, we assume for the moment that $g_i (n)\to \gamma_i \in (0, \infty)$ as $n\to\infty$ for all $i=1,\ldots ,N$.

By irreducibility of $p(i,j)$, up to multiplicative constants there
exists a unique function $\kappa: \{ 1,\ldots, N\} \to (0,\infty)$ such that
for all $i$,
\be\label{harmo}
\sum_{i=1}^N\left(\kappa_j p(j,i)-\kappa_i p(i,j)\right) = 0\ .
\ee
Under these conditions it is well known \cite{andjel} that the zero-range process has stationary product measures with marginals
\be\label{zeromar}
\nu_\phi^i (n) =\frac{1}{z_i(\phi)} \frac{\phi^n \kappa_i^n}{\prod_{k=1}^n g_i (k)}\ ,
\ee
which are of the form \eqref{generalform} with
\[
w_i (n) =\frac{ \gamma_i^n}{\prod_{k=1}^n g_i (k)}\quad\mbox{and}\quad  \lambda_i  = \frac{\kappa_i}{\gamma_i}\ .
\]
So in order to apply the general result,
we need
\[
\frac{\kappa_N}{\gamma_N} > \max_{i=1}^{N-1} \frac{\kappa_i}{\gamma_i}\ ,
\]
and the subexponentiality condition on $w_i$ follows since
\[
\lim_{n\to\infty} \frac{w_i (n+1)}{w_i (n)} = \lim_{n\to\infty} \frac{\gamma_i}{g_i (n+1)} =1\ .
\]

\br
\begin{enumerate}
\item The case $\gamma_i=\infty$ for some $i\not= N$ can be included as well.
In that case, $z_i (\phi) <\infty$ in \eqref{zeromar} for all $\phi >0$, in particular
for $\phi= \phi_c= 1/\lambda_N$. Therefore the result of Theorem \ref{condt}
still holds.
\item If there are more sites $i$ such that $\lambda_i =\lambda_N$, then
a) of Theorem \ref{condt} holds for all $i$ where
$\lambda_i < \lambda_N$. Item b) becomes that all but a finite amount
of mass is concentrated on the sites where $\lambda_i=\lambda_N$.
\end{enumerate}
\er

Note that we make no assumptions on the jump rates of the zero-range process except a regular limiting behaviour, in particular there are no monotonicity assumptions. The latter have been in place in previous work on inhomogeneous zero-range condensation where the $g_i$ are non-decreasing \cite{krug,landim96,ferrarisisko,andjeletal00}, which made it possible to make much stronger statements including also the time evolution of the condensation. In that sense Theorem \ref{condt} is a generalization of previous results regarding only the stationary distribution.

\section{Condensation in homogeneous\\ inclusion processes}

In this section we study condensation in spatially homogeneous systems.
There are two natural situations where Theorem \ref{thstat} leads to spatially homogeneous product measures $\nu_\phi$. If the $p(i,j)$ are symmetric, i.e. $p(i,j)=p(j,i)$ for all $i,j\in\Lambda$ then the reversibility condition (\ref{lrev}) is fulfilled by taking a constant $\lambda_i =1$ for all $i\in\Lambda$ independent of the geometry of the lattice. The same solution holds for
translation invariant, asymmetric processes according to condition (\ref{dstoch}), where
\[
p(i,j)=q(j-i)\quad\mbox{for some }q:\Lambda\to [0,\infty )\mbox{ with bounded support}\ .
\]
In the second case the lattice also has to be translation invariant, such as $\Lambda =\Z^d$ or a finite subset with periodic boundary conditions. The measures $\nu_\phi$ are then not reversible and the system can support a non-zero stationary current of the form
\[
J(\rho )=\rho \left(\frac{m}{2} +\rho\right)\sum_{k\in\Lambda} k\, q(k)\ .
\]

\subsection{Stationary measures}

In both cases discussed above the inclusion process
has a family of homogeneous stationary product measures with marginals
\be\label{sipmar}
\nu^i_\phi (n)= \frac{1}{z(\phi)}\, \phi^n
\frac{\Gamma\left(\frac{m}2+n\right)}{n!\Gamma\left(\frac{m}2\right)}\ ,
\ee
and the normalizing constant
\be\label{sipz}
z(\phi)= \sum_{n=0}^\infty \phi^n
\frac{\Gamma\left(\frac{m}2+n\right)}{n!\Gamma\left(\frac{m}2\right)}
= \left(1-\phi\right)^{-m/2}\ .
\ee
The measures are well defined for all positive $\phi <\phi_c =1$, and the average number of particles per site is given by
\be\label{sipr}
\rho_m (\phi )=\phi\, \partial_\phi \log z(\phi )=\frac{m}2\frac{\phi}{1-\phi}\ .
\ee
Inverting this relation $\phi_m (\rho )=\frac{\rho}{m/2+\rho}$ allows us -- with a slight abuse of notation -- to index the measures by the density,
\be\label{sipmar2}
\nu^{(m)}_\rho (n)= \frac1{z(\phi_m (\rho ))} \left(\frac{\rho}{m/2+\rho}\right)^n
\frac{\Gamma\left(\frac{m}2+n\right)}{n!\Gamma\left(\frac{m}2\right)}\ .
\ee
We also replace the superscript since the marginals are site-independent and we want to stress the dependence on the parameter $m$.
Since the density can take all values between $0$ and $\infty$, we see that for fixed $m>0$ the attraction between the particles is not strong enough and the inclusion process does not exhibit condensation. However, if we increase the relative strength of the attractive part in the generator (\ref{ipgen}) by taking $m$ smaller and smaller at a fixed density $\rho$, a condensation phenomenon occurs in the limit $m\to 0$.

\bt\label{sctheo}
As $m\to 0$, we have for all $\rho >0$
\beq
\nu^{(m)}_\rho (0)&=&\left(\frac{m}{2\rho}\right)^{m/2} \big( 1+o(1)\big) \to 1\qquad\mbox{and}\nonumber\\
\nu^{(m)}_\rho (n)&=&\frac{m}{2}\left(\frac{m}{2\rho}\right)^{m/2} \left( 1-\frac{m}{2\rho}\right)^n n^{m/2-1} \big( 1+o(1)\big)\to 0
\eeq
for $n\geq 1$, which implies
\be\label{nulim}
\frac2{m} \,\nu^{(m)}_\rho (n)\to \frac1{n}\ .
\ee
\et

\bpr
By direct computation we get that
\be\label{zasymp}
z\left(\phi_m (\rho )\right)=\left( 1-\frac{\rho}{m/2+\rho}\right)^{-m/2} =\left(\frac{2\rho}{m} \right)^{m/2} \big( 1+o(1)\big)\to 1
\ee
as $m\to 0$, which directly implies the statement for $\nu^{(m)}_\rho (0)=1/z\left(\phi_m (\rho )\right)$. For every fixed $n\geq 1$ we have
\[
\left(\frac{\rho}{m/2+\rho}\right)^n =\left( 1-\frac{m}{2\rho}\right)^n \big( 1+o(1)\big)\to 1\ ,
\]
and using $\Gamma (x)\sim\frac1{x}$ as $x\to 0$ and (\ref{asweight}) we obtain
\[
\frac{\Gamma\left(\frac{m}2+n\right)}{n!\Gamma\left(\frac{m}2\right)} =\frac{m}2 \, n^{m/2-1} \big( 1+o(1)\big)\to 0\ ,
\]
which implies the second statement. The limit in (\ref{nulim}) follows immediately from the asymptotic behaviour.
\epr

Therefore, for small diffusion rate $m$ sites are either empty with very high probability, or contain a large number of particles to match the fixed expected value $\rho >0$. From theorem \ref{sctheo} we infer the following leading-order behaviour for small fixed $m$,
\be\label{nuasymp}
\nu^{(m)}_\rho (n)\simeq \frac{m}2\left\{\begin{array}{cl} n^{-1}&,\ 1\ll n\ll 2\rho /m\\ (1-\frac{m}{2\rho})^n n^{m/2-1}&,\ n\gg 2\rho /m\end{array}\right.\ ,
\ee
where we have used
\[
\left(1-\frac{m}{2\rho}\right)^n \simeq \left( 1-\frac{mn}{2\rho}\right) \simeq 1\quad\mbox{for }n\ll 2\rho /m\ ,
\]
with the notation $a_m \simeq b_m$ if $a_m /b_m \to 1$ as $m\to 0$.

\begin{figure}%
\centering
\includegraphics[width=0.7\textwidth]{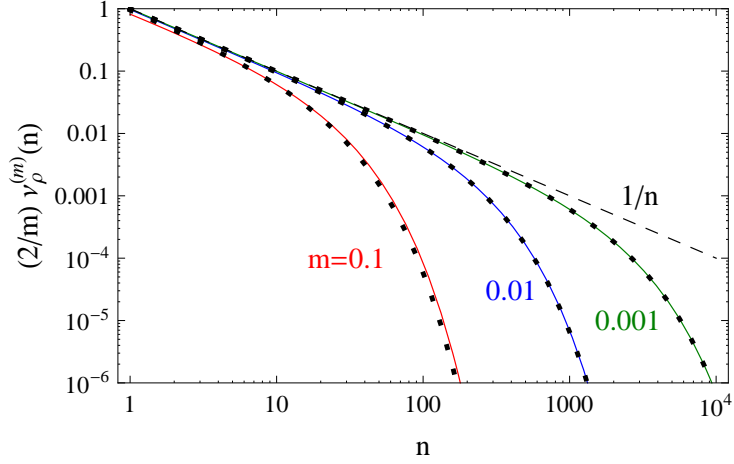}%
\caption{The scaled marginal $(2/m) \nu_\rho^{(m)}$ for $\rho =1$ and for several values of $m$ (full colored lines). The asymptotic behaviour as given in (\ref{nuasymp}) is indicated by dotted lines.}%
\label{fig1}%
\end{figure}

So the marginals show an approximate power law decay with exponent $-1$, until an exponential cut-off sets in at the scale $n\sim 2\rho /m$. This is illustrated in Figure \ref{fig1}, where we see that the asymptotic behaviour for large $n$ fits very well also for smaller values of $n$. Despite the small prefactor $m/2$ the density $\rho >0$ is realized by the asymptotic heavy-tail behaviour, and for each $m>0$ the distribution is normalized due to the cut-off. Conditioned on a site being non-empty, its distribution is given by $\nu^{(m)}_\rho (n)/\big( 1-\nu_\rho^{(m)} (0)\big)$. Using that to leading order
\[
1-\nu_\rho^{(m)} (0)\simeq 1-\left(\frac{m}{2\rho}\right)^{m/2} \simeq -\frac{m}2\log m\ ,
\]
we get with (\ref{nuasymp}) for the conditional distributions
\be\label{degen}
\frac{\nu^{(m)}_\rho (n)}{1-\nu_\rho^{(m)} (0)}\, |\log m|\to \frac1{n}\quad\mbox{as}\quad m\to 0\ .
\ee
Like in (\ref{nulim}), convergence is clearly non-uniform due to the cut-off, and the limit is not a probability distribution. The interpretation of this result in terms of condensation depends on the geometry and is different for finite and infinite lattices $\Lambda$, as discussed below.

\subsection{Finite systems}

For finite lattices one can condition on the total number of particles in the system, defining the canonical measures as in Section 2.2. The basic features of this approach can already be understood on a system with two sites and $\Lambda =\{ 1,2\}$. Let $\eta_1 ,\eta_2$ be two random variables each distributed as $\nu^{(m)}_\rho$ and consider their joint distribution $\mu_m^K$ conditioned on their sum being equal to $K\in\N$, i.e.
\be
\mu_m^K :=\nu_\rho \big(.\,\big|\eta_1 +\eta_2 =K\big)\ .
\ee
For each $K\in\N$ and $m>0$ the inclusion process is irreducible and $\mu_m^K$ is the unique stationary measure (cf.(\ref{mustat})). A first observation is that, as before, $\mu_m^K$ does in fact not depend on $\rho$ since due to cancellation
\beq
\lefteqn{\mu_m^K (\eta_1 =n_1 ,\eta_2 =n_2 )=\frac{\delta_{n_1+n_2,K}\,\nu^{(m)}_\rho (n_1)\nu^{(m)}_\rho (n_2)}{\sum_{l=0}^K \nu^{(m)}_\rho (l)\nu^{(m)}_\rho (K-l)}
}\nonumber\\
& &\quad = \frac{\delta_{n_1 +n_2,K}\,\Gamma(m/2+n_1)\Gamma(m/2+n_2)/(n_1 !n_2 !)}{\sum_{l=0}^K \Gamma(m/2+l)\Gamma(m/2+K-l)/(l!(K-l)!)}\ .
\eeq

\bp
In the limit $m\to 0$ we have for all $K>0$
\be\label{muklim}
\mu_m^K \to \frac12 (\delta_{(K,0)} +\delta_{(0,K)} )\ ,
\ee
i.e. all particles concentrate on one of the sites with equal probability.
\ep
\bpr
With $\eta_2 =K-\eta_1$ we have
\be\label{pr1}
\mu_m^K (\eta_1 =n,\eta_2 =K-n)=\frac{\Gamma(\frac{m}2+n)\Gamma(\frac{m}2+K-n)/(n!(K-n)!)}{\sum_{l=0}^K \Gamma(\frac{m}2{+}l)\Gamma(\frac{m}2{+}K{-}l)/(l!(K-l)!)}\ .
\ee
In the normalizing sum, as $m\to 0$, the two terms for $l=0,K$ diverge like $\Gamma (m/2)/K$, whereas the rest of the sum converges. Also the term in the numerator of $\mu_m^K (\eta_1 =n)$ diverges like $\Gamma (m/2)/K$ if $n=0$ or $K$ and is finite otherwise. This implies the result.
\epr

The interpretation is that as $m\to 0$ aggregation dominates more and more over diffusion and the particles tend to cluster on one of the lattice sites.
The onset of condensation for small $m$ can be well illustrated in the limit of infinitely many particles.

\bp
In the limit $K\to\infty$ we have for all $m>0$,
\be
\left(\frac{\eta_1}{K},\frac{\eta_2}{K}\right)\stackrel{\mu^K}{\longrightarrow} (B,1-B) \qquad\mbox{in distribution}\ ,
\ee
where $B\in [0,1]$ is a continuous random variable with $\mathrm{Beta}\left(\frac{m}{2},\frac{m}{2}\right)$ distribution and PDF
\be\label{betapdf}
f_B (x)=\frac{\Gamma (m/2)^2}{\Gamma (m)}\, x^{m/2-1} (1-x)^{m/2-1} \ ,\quad x\in [0,1]\ .
\ee
\ep

\bpr
Using (\ref{asweight}) we get as $K\to\infty$ and $n/K\to x\in [0,1]$ for the asymptotic form of the numerator of (\ref{pr1})
\[
K^{m-2} x^{m/2-1} (1-x)^{m/2-1}\ .
\]
For the denominator we get the integral
\[
K^{m-2} \int_0^1 y^{m/2-1} (1-y)^{m/2-1} K\, dy=K^{m-1} \frac{\Gamma (m)}{\Gamma (m/2)^2}\ ,
\]
using the representation $B(r,s)=\frac{\Gamma (r+s)}{\Gamma (r)\Gamma (s)}$ for the Beta function. Thus we have that $K\mu_m^K (\eta_1 =n)\to f_B (x)$ converges to the PDF of the Beta$\left(\frac{m}{2},\frac{m}{2}\right)$ distribution.
\epr

\begin{figure}%
\centering
\includegraphics[width=0.7\textwidth]{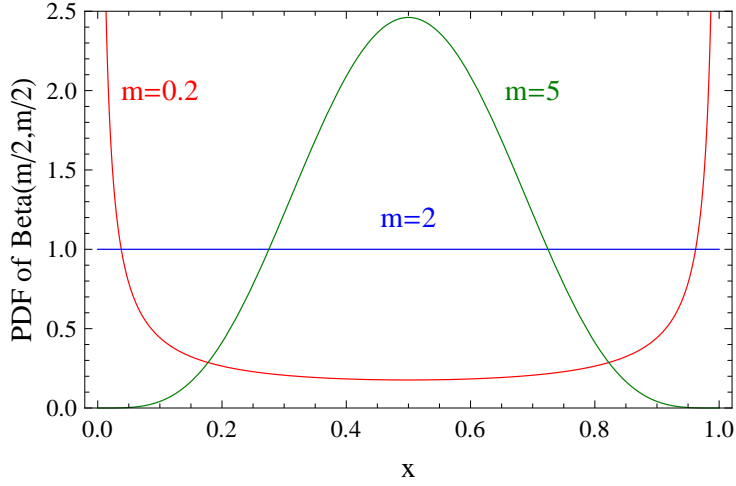}%
\caption{The limit distribution of $\eta_1 /K$ as $K\to\infty$, given by the PDF of Beta$\left(\frac{m}{2},\frac{m}{2}\right)$ (cf. \ref{betapdf}) for several values of $m$.}%
\label{fig2}%
\end{figure}

We see that for $m<2$ one site contains most of the particles while for $m>2$ both sites are likely to have around $K/2$ particles. The boundary case is $m=2$, where the particles are distributed uniformly among the two sites. This is a standard property of the symmetric Beta distribution and is illustrated in Figure \ref{fig2}.
In the limit $m\to 0$ we recover the degenerate distribution (\ref{muklim}).

\br
\begin{itemize}
    \item[a)] The result (\ref{muklim}) can be immediately generalized to a finite set $\Lambda_N =\{ 1,\ldots ,N\}$ of $N\geq 2$ sites. In the limit $m\to 0$ we have for all $K\in\N$
\be\label{nstat}
\mu_m^K \to\frac1{N}\sum_{i=1}^N \delta_{K\, e_i}\ ,
\ee
where $e_i =(..,0,1,0,..)\in\R^N$ is the standard unit vector in direction $i$.
\item[b)] In the absence of diffusion for $m=0$ the inclusion process has in general many absorbing states which exhibit several isolated piles of particles. However, if the $p(i,j)>0$ for all $i,j\in\Lambda_N$, then all absorbing states have exactly one pile containing all the particles. The stationary measures are then all possible mixtures
\be
\sum_{i=1}^N \alpha_i \delta_{K\, e_i}\quad\mbox{with}\quad \alpha_i \in [0,1]\quad\mbox{and}\quad\sum_i \alpha_i =1\ .
\ee
The limit result (\ref{nstat}) leads only to the symmetric mixture, due to homogeneity and ergodicity of the process for $m>0$.
\end{itemize}
\er

\noindent\textbf{Connection to zero-range processes.}\\
This result is slightly different from most previous work on homogeneous zero-range condensation,
which is mostly discussed in the limit of infinitely many particles \cite{sisko} or the thermodynamic limit \cite{grossketal03,evansetal05}. In this case, above a certain density or particle number all sites have heavy-tailed distributions and condensation is a consequence of large deviation properties of such random variables, as discussed in detail in \cite{armendarizetal09}.

For the inclusion process we discuss the two extreme cases of a finite and an infinite lattice (see next section), in the limit of a vanishing system parameter $m\to 0$. The distributions of the occupation numbers always have exponential tails due to the cut-off (\ref{nuasymp}), which disappears in the limit in a non-uniform way. This is very similar to results in \cite{gs08}, where a parameter was varied together with the system size in a joint limit. Analogous results are phrased here in terms of the law of large numbers in the next section. Size-dependent system parameters have also been studied in \cite{schwarzkopf}, which can lead to a cut-off similar to (\ref{nuasymp}) and a typical maximal cluster size also in zero-range processes.

As a further difference to zero-range condensation, there is no non-trivial critical density $\rho_c$ for the distribution of sites outside the maximum in the inclusion process. In fact, in the limit $m\to 0$ all $N$ particles condense on a single site, which corresponds to $\rho_c =0$ and is an absorbing state for the dynamics with $m=0$. This is related to results on zero-range processes where the jump rates vanish in the limit of infinite occupation number, which has been studied in \cite{march} and more recently also in \cite{jeon}.

\subsection{The infinite-volume limit}

For finite systems with a fixed number of particles the exponential part of the product measures that leads to a cut-off for large $n$ (cf. (\ref{nuasymp})) did not play any role due to cancellation, but will be of importance for infinite systems. For simplicity we consider stationary configurations of the symmetric inclusion process (SIP) on the infinite lattice $\Lambda =\N$ which leads to a family of iid random variables $\eta_1 ,\eta_2 ,\ldots$ with distribution $\nu_\rho^{(m)}$ (\ref{sipmar2}). In this context the condensation phenomenon for $m \to 0$ can be formulated as a breakdown of the usual law of large numbers.

For every $m >0$ by definition $\E (\eta_i )=\rho$ and a usual law of large numbers holds, i.e.
\be
S_K :=\frac1{K}\sum_{i=1}^K \eta_i \to\rho\quad a.s.\quad\mbox{as}\quad K\to\infty\ .
\ee
On the other hand, $\eta_i \to 0$ as $m\to 0$ in distribution, and therefore we have for all $K\in\N$ even for the unnormalized sums
\[
\sum_{i=1}^K \eta_i \to 0\quad\mbox{in distr.\quad as}\quad m\to 0\ .
\]
This implies that the limiting behaviour of the empirical mean as $K\to\infty$ and $m\to 0$ depends on the order of limits. Thus we are interested in the joint limit $K_m \to\infty$ as $m\to 0$ to identify the scale on which the law of large numbers changes behaviour. It turns out that there are two interesting scales for $K_m$.

\bp
Let $\kappa_m =\frac{-1}{m\log m}$. Then as $m\to 0$ we have (in distr.)
\be\label{clim}
\Delta_{K_m} :=\sum_{i=1}^{K_m} \big( 1-\delta_{0,\eta_i}\big) \longrightarrow\left\{\begin{array}{cl} 0 &,\ K_m \ll\kappa_m\\
W_\delta &,\ K_m /\kappa_m \to\delta\in (0,\infty )\\
\infty &,\ K_m \gg\kappa_m\end{array}\right.\ ,
\ee
where $W_\delta \sim Poi(\delta /2)$ is a Poisson random variable with mean $\delta /2$. In the last case,\quad $\Delta_{K_m} =\frac{K_m}{2\kappa_m}\big( 1+o(1)\big)$\ .\\
Furthermore, on the larger scale $1/m\gg\kappa_m$ we have
\be\label{elim}
S_{K_m} =\frac1{K_m}\sum_{i=1}^{K_m} \eta_i \longrightarrow\left\{\begin{array}{cl} 0 &,\ K_m \ll 1/m\\
X_\gamma &,\ K_m m \to\gamma\in (0,\infty )\\
\rho &,\ K_m \gg 1/m\end{array}\right.\ ,
\ee
where $X_\gamma \sim \mathrm{Gamma}\big(\frac{\gamma}2,\frac{2\rho}{\gamma}\big)$ is a Gamma random variable with mean $\rho$.
\ep

\bpr
Denote the probability of $\eta_i >0$ by
\be\label{pasy}
p_m :=1-\nu_\rho^{(m)} (0)=-\frac{m}{2}\log m \big( 1+o(1)\big) =\frac1{2\kappa_m} \big( 1+o(1)\big)\ ,
\ee
with asymptotics for $m\to 0$. Then $1-\delta_{0,\eta_i}\sim \mathrm{Be}(p_m )$ are i.i.d.\ Bernoulli random variables and therefore $\Delta_{K_m} \sim \mathrm{Bi}(K_m ,p_m )$ is a Binomial with
\[
\Pr (\Delta_{K_m} =n)={K_m \choose n} p_m^n (1-p_m )^{K_m -n}\ ,
\]
counting the non-zero contributions to the sum $S_{K_m}$.
$p_m \to 0$ as $m\to 0$ with asymptotics given in (\ref{pasy}), and (\ref{clim}) is a well-known scaling result for Binomial r.v.s. Since the rescaled random variables $(1-\delta_{0,\eta_i} )/p_m$ have mean $1$, we have by the ususal law of large numbers
\[
\frac{\Delta_{K_m}}{K_m p_m}=\frac1{K_m} \sum_{i=1}^{K_m} \frac1{p_m}(1-\delta_{0,\eta_i} )\to 1\ .
\]
This holds whenever $K_m p_m \to\infty$ or, equivalently, $K_m \gg\kappa_m$ since the sum will have infinitely many non-zero contributions, and implies that $\Delta_{K_m} =\frac{K_m}{2\kappa_m}\big( 1+o(1)\big)$.

Analogous to (\ref{sipz}) we get for the characteristic function of $\eta_i$
\[
\chi_\eta (t)=\E \big( e^{it\eta_1}\big) = \left(\frac{1-\phi}{1-e^{it}\phi}\right)^{m/2}\ .
\]
For the rescaled sum $S_{K_m}$ of $K_m$ independent r.v.s we get
\[
\chi_S (t)=\chi_\eta (t/K_m)^{K_m} =\left( 1+\frac{2\rho}{m} \left( 1-e^{it/K_m}\right)\right)^{-K_m m/2}\ ,
\]
where we used $\rho =\rho_m (\phi )=\frac{m}{2}\frac{\phi}{1-\phi}$ as in (\ref{sipr}) to fix the density. As $K\to\infty$ we have for all complex $z\neq 0$
\be\label{logk}
1-z^{1/K} =-\frac1{K}\log z\,\big( 1+o(1)\big)\ .
\ee
This leads to the asymptotics
\[
\chi_S (t)=\left( 1-\frac{2\rho}{K_m m}\, it\right)^{-K_m m/2} \big( 1+o(1)\big)\ ,
\]
since the correction terms from (\ref{logk}) are of order $1/K_m \ll 1/(mK_m)$.
Therefore, as $m\to 0$
\[
\chi_S (t)\longrightarrow\left\{\begin{array}{cl} 1&,\ mK_m \to 0\\ e^{it\rho} &,\ mK_m \to\infty \end{array}\right.\ ,
\]
which implies the weak law of large numbers in the two extreme cases of (\ref{elim}). In the intermediate case $mK_m \to\gamma$ we have
\[
\chi_S (t)\to\left( 1-\frac{2\rho}{\gamma}\, it\right)^{-\gamma /2}\ ,
\]
which is the characteristic function of a Gamma$\big(\frac{\gamma}2,\frac{2\rho}{\gamma}\big)$ random variable.
\epr

\begin{figure}%
\centering
\includegraphics[width=0.7\textwidth]{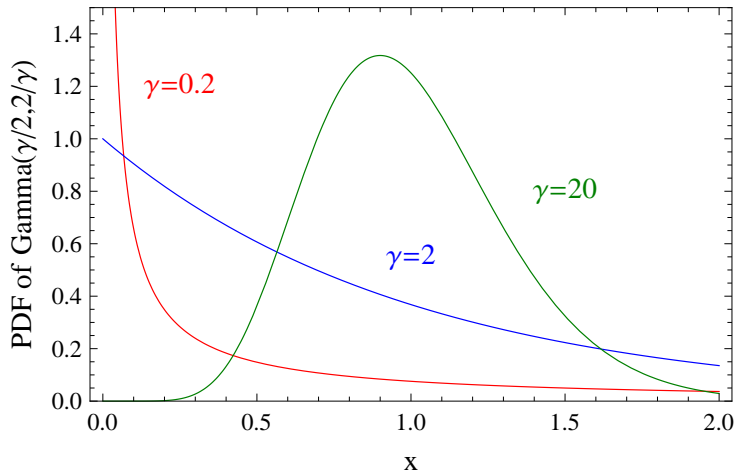}%
\caption{The limit distribution of $S_{K_m}$ as $m\to 0$ with $m K_m\to\gamma$, given by the PDF of a Gamma$\big(\frac{\gamma}2,\frac{2\rho}{\gamma}\big)$ random variable (\ref{elim}). In all cases $\rho =1$, and increasing $\gamma$ (3 values shown) interpolates between the deterministic limits $0$ and $\rho$.}%
\label{fig3}%
\end{figure}

This result leads to the following interpretation for the limiting behaviour of $S_{K_m}$ as $m\to 0$.
\begin{itemize}
\item[a)] $K_m \ll\kappa _m$: There are no non-zero contributions to $S_{K_m}$ and even the unnormalized sum
$K_m S_{K_m} \to 0$.
\item[b)] $K_m \sim\kappa_m$: There is a finite (Poisson distributed) number of non-zero contributions to $S_{K_m}$,
but still $S_{K_m} \to 0$. Since the law of these contributions becomes degenerate as $m\to 0$ (cf. (\ref{degen})) we have no scaling law for $K_m S_{K_m}$.
\item[c)]
$\kappa_m \ll K_m\ll 1/m$: $S_{K_m}$ has an infinite number of non-zero contributions, but still vanishes as $m\to 0$.
\item[d)]
$K_m \sim 1/m$: $S_{K_m}$ has a random limiting value (Gamma distributed) with mean $\rho$, and infinitely many non-zero contributions. This interpolates between the deterministic limits $0$ and $\rho$, as shown in Fig.\ \ref{fig3}.
\item[e)]
$K_m \gg 1/m$: The usual weak law of large numbers holds, i.e. $S_{K_m} \to\rho$ as $m\to 0$.
\end{itemize}

If we interpret $\eta_1 ,\eta_2 ,\ldots$ as a configuration of the inclusion process, this result gives detailed information about the structure of such configurations as $m\to 0$. They are in direct analogy to results in \cite{gs08} on a particular zero-range process, which have just been formulated in an inverted fashion corresponding to a parameter $m_K \to 0$ in the limit $K\to\infty$.

\section{The Brownian energy process}
In \cite{gkrv} we introduced the Brownian energy process with parameter $m>0$ (abbreviation BEP(m)), and
explained how, for integer values of $m$ it is related to the Brownian momentum
process with $m$ momenta per site.

More precisely, the BEP(m) is an interacting diffusion process on $\Omega_N=[0,\infty)^{1,\ldots,N}$
with generator
\beq\label{bepgen}
L f(x) &=& \sum_{i=1}^{N-1} 4x_i x_{i+1}
\left(\frac{\partial}{\partial x_i}-\frac{\partial}{\partial x_{i+1}}\right)^2
\nonumber\\
&& \qquad -2m (x_i-x_{i+1})\left(\frac{\partial}{\partial x_i}-\frac{\partial}{\partial x_{i+1}}\right)\ ,
\eeq
where for a configuration of ``energies'' $x\in \Omega_N$, $x_i$ denotes
the energy at site $i\in \{1,\ldots, N\}$.

In \cite{gkr} we introduced an asymmetric version of the Brownian momentum
process. This model was later studied in \cite{ber}. Motivated by this asymmetric modification
of the Brownian momentum process, we now introduce an asymmetric version of BEP(m)
via its generator
\beq\label{abepgen}
L f(x) &=& \sum_{i=1}^{N-1} 4x_i x_{i+1}
\left(\frac{\partial}{\partial x_i}-\frac{\partial}{\partial x_{i+1}}\right)^2
\nonumber\\
&& -2m (x_i-x_{i+1})\left(\frac{\partial}{\partial x_i}-\frac{\partial}{\partial x_{i+1}}\right)
\nonumber\\
&& -2E x_i x_{i+1} \left(\frac{\partial}{\partial x_i}-\frac{\partial}{\partial x_{i+1}}\right)\ .
\eeq
We focus on a one-dimensional nearest-neighbour lattice as for the ASIP in Section 3, but the definition could of course be generalized to arbitrary geometries.
Obviously, the total energy $f(x)=\sum_{i=1}^N x_i$ is conserved, and for $E>0$ the process has a drift to the left, which can most easily be seen from the stationary measures discussed in the next section.

\subsection{Condensation in the ABEP}

We first consider $m=2$, $E>0$, and two sites. This is the simplest
case because the marginals of the stationary distribution are exponential, which makes explicit computations simple. The generalization to $m>0$ and more sites is easy.

The generator, written in the variables $(x_1,x_2)=: (u,v)$, then reads:
\begin{eqnarray*}
L&=& 4uv\left(\frac{\partial}{\partial u}-\frac{\partial}{\partial v}\right)^2
\nonumber\\
&& - 4 (u-v)\left(\frac{\partial}{\partial u}-\frac{\partial}{\partial v}\right)
\nonumber\\
&& +2Euv \left(\frac{\partial}{\partial u}-\frac{\partial}{\partial v}\right)
\end{eqnarray*}
The adjoint (in $L^2 (\R, dx)$) is given by (the closure of the operator)
\begin{eqnarray*}
L^* &=& 4uv \left(\frac{\partial}{\partial u}-\frac{\partial}{\partial v}\right)^2
\nonumber\\
&&-4(u-v) \left(\frac{\partial}{\partial u}-\frac{\partial}{\partial v}\right)
\nonumber\\
&&+2E (u-v) - 2Euv \left(\frac{\partial}{\partial u}-\frac{\partial}{\partial v}\right)\ .
\end{eqnarray*}
As an ansatz for the density of the stationary distribution we put
\be\label{expan}
f(u,v) = abe^{-au} e^{-bv}
\ee
with $a,b >0$. Plugging this in the equation for the stationary density
$L^* f=0$ gives
\[
4uv (b-a)^2 + 4 (v-u) (b-a) + 2E (u-v) - 2E uv (b-a)=0\ ,
\]
which leads to
\be\label{relE}
b= a + \frac{E}{2}
\ee
and
\be\label{statdist}
f(u,v)= a(a+E/2)e^{-au} e^{-av} e^{-Ev/2} \ .
\ee
In order to state our condensation result, denote
by $(U_K, V_K)$ the pair $(U,V)$ with probability density \eqref{statdist} conditioned
on $U+V=K$. We then have the following result, which should be thought
of as the analogue of Theorem \ref{condt}, but now in continuous state space setting.
\bt
\begin{itemize}
\item[a)] As $K\to\infty$, $V_K$ converges in distribution to a random variable
with exponential distribution with parameter $E/2$, i.e.,
with probability density $(E/2) e^{-u(E/2)}$.
\item[b)] As $K\to\infty$, $U_K/K\to 1$ almost surely.
\end{itemize}
\et
\bpr
The proof is a direct computation.
Put $\lambda = a+ E/2$, $\lambda' =a$, then $\lambda >\lambda'$, $\lambda-\lambda' =E/2$.
First note that the distribution of $U+V$ has probabily density
$\frac{\lambda\lambda'}{\lambda-\lambda'} (e^{-\lambda 'x}- e^{-\lambda x})$.

Next, the conditional density of $V$ given $U+V=K$ is given by
\[
\frac{\lambda\lambda' (\lambda-\lambda') e^{-\lambda u} e^{-\lambda' (K-u)}}
{\lambda\lambda' (e^{-\lambda' K} - e^{-\lambda K})} =
\frac{(\lambda-\lambda')e^{-(\lambda-\lambda') u}}{1-e^{-(\lambda-\lambda')K}}
\]
which converges, as $K\to\infty$ to $(\lambda-\lambda')e^{-(\lambda-\lambda') u}$,
implying statement a) of the theorem.
To prove statement b): choose $0<\delta<1$, then
\begin{eqnarray*}
\lefteqn{\Pr\big( U \leq (1-\delta)K \big| U+V=K\big)
=
\frac{\int_0^{ (1-\delta)K} \lambda\lambda' (\lambda-\lambda') e^{-x\lambda'}e^{-(K-x)\lambda} dx}
{\lambda\lambda' (e^{-\lambda' K } -e^{-\lambda K})}}
\nonumber\\
& &\quad =
\frac{(\lambda-\lambda')\int_0^{ (1-\delta)K} e^{-x(\lambda-\lambda')} dx}
{e^{(\lambda-\lambda') K } - 1}
=\frac{e^{(\lambda-\lambda') K(1-\delta) } - 1}{e^{(\lambda-\lambda') K } - 1} \qquad\qquad\mbox{}\nonumber\\
& &
\quad =\big( e^{-\delta (\lambda-\lambda')}\big)^K \frac{1-e^{(\lambda'-\lambda) K}}{1-e^{(\lambda'-\lambda) K }}\to 0
\end{eqnarray*}
as $K\to\infty$. As in the proof of Theorem \ref{condt} the bound is summable in $K$ if we choose $\delta =1/\sqrt{K}$ and $U_K /K\leq 1$ by definition, which implies almost sure convergence.
\epr

To generalize the previous computation to the case of $N$ sites and
general parameter $m>0$, it is easy to check along the lines of the proof of Theorem \ref{thstat} that the process with generator \eqref{abepgen} has
a stationary measure which is a product of Gamma distributions
with identical shape parameter $m$ and site-dependent location parameter.
More precisely, the PDF is given by
\be\label{abepdensit}
f(x_1,\ldots,x_N) = \prod_{i=1}^N \frac{a_i^{m/2}x_i^{m/2-1} e^{-a_i x_i}}{ \Gamma (m/2)}
\ee
with
\be\label{beei}
a_i = a + \frac{(i-1)E}{2}
\ee
for $i\in \{ 1,\ldots, N\}$.
After conditioning on the sum $X_1+\ldots +X_N= K$
we find, again by simple
explicit computation, in the limit $K\to\infty$ that
$X_1/K$ converges to $1$ almost surely, and that
for $i=2,\ldots, N$, the law of $X_i$ converges to a shifted Gamma distribution
with density
\be\label{shiftgam}
\lim_{K\to\infty}f_{X_i|X_1+\ldots X_N=K} (x_i)= C_i x_i^{\frac{m}{2}-1} e^{-\frac{(i-1)Ex_i}{2}} \ 
\ee
where $C_i = (\frac{i-1}{2} e )^{\frac{m}{2}}/\Gamma(\frac{m}{2})$ is a normalization constant.

The interpretation of this result is the same as in the discrete case for the ASIP. Here almost all energy concentrates on the lattice site with the heaviest tail in the stationary distribution.

\subsection{Generalizations}

Exactly as in the case of condensation in the ASIP (section 3.2), we can formulate
a more general condensation result for independent random variables $X_1,\ldots, X_N$
with values in $[0, \infty)$ and marginal densities
\be\label{cmargis}
f_{X_i} (x) = \frac{1}{z_i (\mu)}\ e^{-\lambda_i x} w_i (x)\, e^{\mu x}\ .
\ee
where $0<\lambda_1<\min_{j=2}^N\lambda_j$. Here a notation with so-called chemical potentials $\mu\in\R$ is more convenient than the fugacity variable $\phi =e^\mu$ used for the SIP, and values $-\infty <\mu<\mu_c:=\lambda_1$ are possible. The normalization
\[
z_i (\mu )=\int_0^\infty e^{-\lambda_i x} w_i (x)\, e^{\mu x}\, dx
\]
is finite for $\mu <\mu_c$, and for indices $i<N$ also $z_i (\mu_c )<\infty$. The $w_i :[0, \infty)\to [0, \infty)$ are subexponential in the sense that for all
$y\in \R$
\be\label{suby}
\lim_{x\to\infty} \frac{w(x+y)}{w(x)} =1\ .
\ee
The proof of this result follows
the same steps as the proof of the analogous discrete result,
except that we have to replace sums by integrals. As this is a straightforward
extension, we leave the proof to the reader.

\bt\label{gencondcont}
Denote by $(Y^K_1, \ldots, Y^K_N)$ the random variables
$(X_1,\ldots, X_N)$ conditioned on $X_1+\ldots +X_N=K$.
Then under the above conditions (\ref{cmargis}) and (\ref{suby}) we have as $K\to\infty$:
\begin{itemize}
\item[a)]
Condensation on the site with the heaviest tail, i.e.
\[
\frac{Y^K_1}{K}\to 1\quad\mbox{almost surely}\ ;
\]
\item[b)]
Convergence to the critical distribution with $\mu =\mu_c$ for other sites, i.e.
\[
(Y^K_2 ,\ldots ,Y^K_N )\to (Y_2 ,\ldots ,Y_N ) \quad\mbox{in distribution}\ ,
\]
where the $Y_i$ are independent with densities
\[
f_{Y_i} (y)= \frac{1}{z_i (\mu_c)}e^{-\lambda_i y} e^{\mu_c y} w_i (y)\ .
\]
\end{itemize}
\et

\br
In the limit $m\to 0$, also for spatially homogeneous Brownian
energy processes there will be a condensation phenomenon as $m\to 0$
completely analogous to the results in Section 4 for the inclusion process.
Indeed, for a fixed average energy $\rho >0$ (taking $a_i =m/(2\rho)$ in (\ref{abepdensit})), the marginal densities of the stationary product measure are
\[
f_{X_i} (x_i)=\frac1{\Gamma (m/2)}\left(\frac{m}{2\rho}\right)^{m/2}x_i^{m/2-1} e^{-mx_i /(2\rho )}\ .
\]
Analogous to Theorem \ref{sctheo} one can easily show that this implies
\[
\Pr (X_i <\delta )=\int_0^\delta f_{X_i} (x_i)\, dx_i \to 1
\]
for all $\delta >0$ as $m\to 0$, so that $X_i \to 0$ in probability. Further, all statements following from Theorem \ref{sctheo} in Section 4 can be derived in an appropriate version for continuous variables.
\er

\section{Conclusion}

We have studied condensation phenomena for random variables with exponential tails, which arise in the inclusion process and related particle systems.
In general, condensation can be due to the presence of subexponential tails resulting from a strong particle attraction, which has been studied in detail in the context of zero-range processes \cite{evans00,grossketal03,godreche,sisko,armendarizetal09,beltran}.
For exponential tails considered in this work, the attraction between particles alone is not strong enough and a second ingredient is needed for condensation.

One possibility are spatial inhomogeneities, which will lead to a non-zero fraction of the particles to cluster on the sites with the heaviest tails in the limit of infinitely many particles. Our result on this in Section 3 applies in great generality, extending also previous related work on zero-range process \cite{krug,landim96,andjeletal00,ferrarisisko}.
For homogeneous systems, varying a system parameter can induce condensation for fixed total particle density as studied in Section 4 for the inclusion process. Previous results in that direction include \cite{gs08,schwarzkopf} for zero-range processes and also \cite{zielen} for a continuous mass model.
The Brownian energy process studied in Section 5 provides an interesting example where both versions of condensation can be studied in a system with continuous state space and dynamics. Condensation for continuous variables has been studied before in the random average process \cite{zielen} and mass transport models \cite{zia,zia2}, all of which use a discontinuous redistribution of mass (or energy) following a jump process.

To summarize, inclusion processes and related systems such as the BEP provide a rich class of models that exhibit condensation phenomena of several kinds in the presence of exponential tails, the description of which applies also in more general situations.
For inhomogeneous models we have focused on finite systems, and a further question would be to consider thermodynamic limits where, for example, inhomogeneity is due to random disorder as studied in \cite{krug,landim96,andjeletal00,ferrarisisko} for zero-range processes. In the homogeneous case it would be of great interest to exploit duality in the SIP and BEP to get results on
{\em the dynamics} of condensation.

\section*{Acknowledgements}
SG is grateful for the support and hospitality of the Hausdorff Research Institute for Mathematics in Bonn, where part of this project was carried out.


\begin{thebibliography}{99}

\bibitem{gkrv}
C. Giardina, J. Kurchan, F. Redig, K. Vafayi: Duality and hidden symmetries
in interacting particle systems, J. Stat. Phys. \textbf{135}, 25-55 (2009).
\bibitem{grv}
C. Giardina, F. Redig, K. Vafayi: Correlation inequalities for interacting particle
systems with duality, J. Stat. Phys. \textbf{141}, 242-263 (2010).
\bibitem{gkr}
C. Giardina, J. Kurchan, F. Redig: Duality and exact correlations for a model
of heat conduction. J. Math. Phys. \textbf{48}, 033301 (2007).
\bibitem{ber}
C. Bernardin: Superdiffusivity of asymmetric energy model in dimensions 1 and
2. J. Math. Phys. \textbf{49}, 103301 (2008).
\bibitem{evans00}
M.R. Evans: Phase Transitions in One-Dimensional Nonequilibrium Systems. Braz.
J. Phys. \textbf{30}(1), 42-57 (2000).
\bibitem{evansetal05}
M.R. Evans, T. Hanney: Nonequilibrium statistical mechanics of the zero-range process and related models. J. Phys. A: Math. Gen. \textbf{38} R195-R240 (2005).
\bibitem{krug}
J. Krug and P.A. Ferrari. Phase transitions in driven diffusive systems with random rates. J.\ Phys.\ A: Math.\ Gen.\ \textbf{29}, L465-L471 (1996).
\bibitem{landim96}
C. Landim: Hydrodynamic limit for space inhomogeneous one-dimensional totally asymmetric zero-range processes. Ann. Prob. \textbf{24}(2), 599-638 (1996).
\bibitem{andjeletal00}
E.D. Andjel, P.A. Ferrari, H. Guiol, C. Landim: Convergence to the maximal invariant measure for a zero-range process with random rates. Stoch. Proc. Appl. \textbf{90}, 67-81 (2000).
\bibitem{ferrarisisko}
P.A. Ferrari, V. Sisko: Escape of mass in zero-range processes with random rates. in IMS Lecture notes, Asymptotics: Particles, Processes and Inverse Problems. \textbf{55}, 108-120 (2007).
\bibitem{grossketal03}
S. Grosskinsky, G.M. Sch\"utz, H. Spohn:
Condensation in the zero range process: stationary and dynamical properties
J. Stat. Phys. \textbf{113}(3/4), 389-410 (2003).
\bibitem{godreche}
C. Godreche, J.M. Luck: Dynamics of the condensate in zero-range processes. J. Phys. A \textbf{38}, 72157237 (2005).
\bibitem{sisko}
P.A. Ferrari, C. Landim, V. Sisko: Condensation for a fixed number of independent random variables. J. Stat. Phys \textbf{128}(5), 1153-1158 (2007).
\bibitem{armendarizetal09}
I. Armend\'ariz, M. Loulakis: Thermodynamic limit for the invariant measures in supercritical zero range processes. Prob. Theory Rel. Fields \textbf{145}(1-2), 175-188 (2009).
\bibitem{beltran}
J. Beltran, C. Landim: Metastability of reversible condensed zero range processes on a finite set. published online in Probab. Theory Relat. Fields, DOI 10.1007/s00440-010-0337-0 (2011)
\bibitem{liggett}
T.M. Liggett: Interacting Particle Systems, Springer (1985)
\bibitem{andjel}
E.D. Andjel: Invariant measures for the zero range process. Ann. Probability \textbf{10}(3), 525-547 (1982).
\bibitem{gs08}
S. Grosskinsky, G.M. Sch\"utz: Discontinuous condensation transition and nonequivalence of ensembles in a zero-range process. J. Stat. Phys. \textbf{132}(1), 77-108 (2008).
\bibitem{schwarzkopf}
Y. Schwarzkopf, M.R. Evans, D. Mukamel: Zero-range processes with multiple condensates: statics and dynamics. J. Phys. A: Math. Theor. \textbf{41}, 205001 (2008).
\bibitem{march}
I. Jeon, P. March, B. Pittel: Size of the largest cluster under zero-range invariant measures. Ann. Probab. \textbf{28}(3), 1162-1194 (2000).
\bibitem{jeon}
I. Jeon: Phase transition for perfect condensation and instability under the perturbations on jump rates of the zero-range process. J. Phys. A: Math. Theor. \textbf{43}, 235002 (2010).
\bibitem{zielen}
F. Zielen, A. Schadschneider: Broken Ergodicity in a Stochastic Model with Condensation. Phys. Rev. Lett. \textbf{89}, 090601 (2002).
\bibitem{zia}
S.N. Majumdar, M.R. Evans, R.K.P. Zia: The Nature of the Condensate in Mass Transport Models. Phys. Rev. Lett. \textbf{94}, 180601 (2005).
\bibitem{zia2}
M.R. Evans, S.N. Majumdar, R.K.P. Zia: Factorised Steady States in Mass Transport Models on an Arbitrary Graph. J. Phys. A: Math. Gen \textbf{39}, 4859-4873 (2006).

\end{thebibliography}
\end{document}